\documentclass{aa}  

\usepackage{graphicx}
\usepackage{txfonts}
\usepackage{hyperref}
\usepackage{xcolor}
\usepackage{ulem}
\usepackage[flushleft]{threeparttable}
\usepackage{CJK}

\def\arcsecs{{$^{\prime\prime}$}}

\begin{document}

   \title{On the 5-Minute Oscillations of Photospheric and Chromospheric Swirls}

   \author{Jiajia Liu
          \inst{1,2,3},
          David Jess
          \inst{2,4},
          Robert Erd{\'e}lyi
          \inst{5,6,7}
          \and
          Mihalis Mathioudakis\inst{2}
          }

   \institute{Deep Space Exploration Laboratory/School of Earth and Space Sciences, University of Science and Technology of China, Hefei 230026, China
              \and
              Astrophysics Research Centre, School of Mathematics and Physics, Queen's University Belfast, Belfast BT7 1NN, UK
              \and
              CAS Key Laboratory of Geospace Environment, Department of Geophysics and Planetary Sciences, University of Science and Technology of China, Hefei, 230026, China\\
              \email{jiajialiu@ustc.edu.cn}
         \and
             Department of Physics and Astronomy, California State University Northridge, 18111 Nordhoff Street, Northridge, CA 91330, USA
             \and
             Solar Physics and Space Plasma Research Centre (SP2RC),  School of Mathematics and Statistics, University of Sheffield, Hicks Building, Hounsfield Road, Sheffield, S3 7RH, UK
             \and
             Department of Astronomy, E\"otv\"os Lor\'and University, P\'azm\'any P. s\'et\'any 1/A, Budapest, H-1117, Hungary
             \and
             Gyula Bay Zolt\'an Solar Observatory (GSO), Hungarian Solar Physics Foundation (HSPF), Pet\H{o}fi t\'er 3., Gyula, H-5700, Hungary
             }

   \date{Received xx xx, 2022; accepted xx xx , 2023}

 
  \abstract
   {Swirls are ubiquitous in the solar atmosphere. They are believed to be related to the excitation of different modes of magnetohydrodynamic waves and pulses, as well as spicules. However, statistical studies of their collective behaviour are rare.}
   {In this paper, we aim to study the collective, as well as the behaviour of individual photospheric and chromospheric swirls detected by the automated swirl detection algorithm (ASDA) from observations obtained by the Swedish 1-m Solar Telescope and the Hinode satellite.}
   {Detailed analysis of six different parameters of photospheric and chromospheric swirls is performed employing the wavelet analysis. Two clusters of periods with significant wavelet power, one from $3-8$~minutes and the other from $10-14$~minutes, have been found. The former coincides with the dominant period of the global $p$-mode spectrum. Wavelet and Fast Fourier Transform (FFT) analysis of example swirls also reveals similar periods.}
   {These results suggest that global $p$-modes might be important for triggering photospheric and thus chromospheric swirls. A novel scenario of global $p$-modes providing energy and mass fluxes to the upper solar atmosphere via generating swirls, Alfv{\'e}n pulses and spicules is then proposed.}
   {}

   \titlerunning{Oscillation characteristics of swirls}
    \authorrunning{Liu et al.}
   \maketitle
%
\section{Introduction} \label{sec:intro}

Swirls are widely observed in different layers of the solar atmosphere \citep[e.g.,][]{Wang1995, Bonet2008, Attie2009, Balmaceda2010, Wedemeyer2009, Bonet2010, Wedemeyer2012, Su2014, Kato2017, Liu2019ApJ, Liu2019NC, Shetye2019}. Besides their ubiquity, they have been suggested to be related to a number of different magnetohydrodynamic (MHD) processes from the photosphere through to the corona, including MHD waves and pulses \citep[e.g.,][]{Carlsson2009, Fedun2011, Shelyag2013, Shukla2013, Mumford2015, Leonard2018, Murawski2018, Kohutova2020, Battaglia2021}, spicules or jets \citep[e.g.,][]{Kitiashvili2012, Kitiashvili2013, Liu2019NC, Oxley2020, Scalisi2021, Scalisi2021b, 2022ApJ...930..129B, Dey2022}, and (magnetic) bright points or magnetic concentrations \citep[e.g.,][]{Balmaceda2010, 2010ApJ...719L.134J, Liu2019AA, Shetye2019, Murabito2020}. Further observations and numerical simulations suggest that swirls might be able to channel enough energy from the lower to the upper solar atmosphere \citep[e.g.,][]{Wedemeyer2009, Liu2019NC, Yadav2021, Battaglia2021}.

Detailed studies of individual swirls, or a small collection of swirls, have shown intriguing and important properties. For example, \citet{Wedemeyer2009} analysed 10 clear small-scale swirling candidates observed in the chromosphere by the Swedish 1-m Solar Telescope \citep[SST;][]{Scharmer2003} and found that these swirls could be related to the so-called magnetic tornadoes channelling energy into the solar corona. A series of investigations were conducted to establish the properties, including its associated waves and oscillations, of a persistent quiet-Sun swirl that had a lifetime exceeding 1.7~hours \citep{Tziotziou2018,Tziotziou2019,Tziotziou2020}. However, owing to their ubiquity \citep[$>10^5$ swirls in the photosphere at any time, e.g.,][]{Liu2019ApJ} and small scale (with an average radius of several hundred kilometres), statistical studies of manually selected swirls have proven difficult and may introduce unwanted human bias. Several automated detection methods have recently been developed. \citet{Kato2017} presented two detection algorithms both based on a line integral convolution (LIC) imaging technique with one using enhanced vorticity and the other the vorticity strength to identify swirls. Their methods have been tested on a simulated chromosphere generated by the CO$^{5}$BOLD \citep{Freytag12} numerical MHD code. Another method, recently proposed by \cite{Dakanalis2021}, employs a series of processes including image pre-processing, tracing of curved structures, segmentation and clustering, with the last two processes being widely used in machine learning techniques. This method was tested with both synthetic data and observations obtained by the SST and was further suggested to be applicable for quasi-linear fibrillar structure detections following some modifications.

Employing the velocity field information estimated from successive images by Fourier local correlation tracking \citep[FLCT;][]{Welsch2004, Fisher2008}, \citet{Liu2019ApJ} developed an automated swirl detection algorithm (ASDA\footnote{\href{https://github.com/PyDL/ASDA}{https://github.com/PyDL/ASDA}}). ASDA was applied to both photospheric and chromospheric observations \citep{Liu2019ApJ, Liu2019NC} acquired by the Solar Optical Telescope \citep[SOT;][]{Tsuneta2008} on board Hinode \citep{Kosugi2007}, and the CRisp Imaging SpectroPolarimeter \citep[CRISP;][]{Scharmer2006} on the SST \citep{Scharmer2003}. A total number of more than $10^5$ swirls were found in the photosphere at any moment of time, with an average radius of $\sim300$~km, rotating speed of $\sim1$~km{\,}s$^{-1}$, and a lifetime of around 20~s. Correlation analysis between photospheric and chromospheric swirls, together with three-dimensional MHD numerical simulations, suggested that ubiquitous Alfv{\'e}n pulses could be excited by photospheric swirls and travel to the chromosphere \citep{Liu2019NC}. The co-existence of intensity swirls and magnetic swirls \citep{Liu2019AA} in the simulated photosphere generated by the Bifrost code \citep{Gudiksen2011, Carlsson2016} suggested that the necessary condition for the generation of Alfv{\'e}n pulses in the solar atmosphere might be fulfilled. This scenario was confirmed by the MHD numerical simulation using the radiative MHD code CO$^5$BOLD \citep{Battaglia2021}. Recent advances in analytical and numerical simulations \citep{Oxley2020, Scalisi2021, Scalisi2021b, Singh2022} suggest that these Alfv{\'e}n pulses could further drive upward mass motions, e.g., spicules, that propagate in the upper solar atmosphere.

Besides these recent advances in observations, numerical simulations and theories of solar atmospheric swirls, we are not aware of many works that study the collective behaviours of swirls and their relationship with global phenomena of the Sun, such as the 5-minute global acoustic oscillations and the solar activity cycle. In this paper, we present evidence of the 5-minute oscillation of photospheric and chromospheric swirls. This paper is organised as follows: data and methods will be briefly introduced in Sect.~\ref{sec:dam}, before we present results in Sect.~\ref{sec:result} and draw the conclusions in Sect.~\ref{sec:conc}.

\section{Data and Methods} \label{sec:dam}

\begin{figure*}[ht!]
\includegraphics[width=\textwidth]{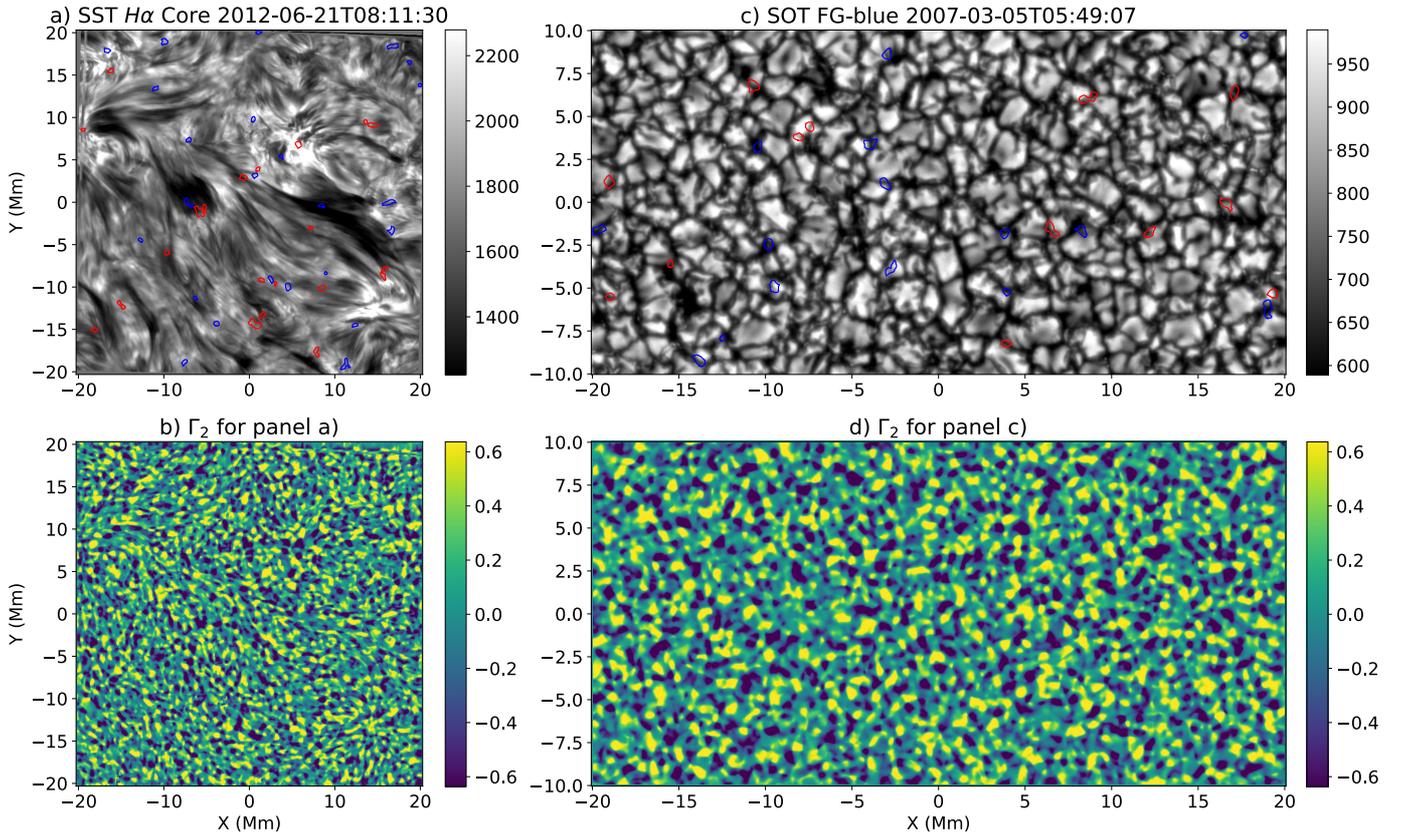}
\caption{\textbf{Examples of swirls detected from SST and Hinode Observations.} Black and white backgrounds in panels a) and c) are the SST H$\alpha$ line core chromospheric and SOT FG-blue photospheric observations on 21$^{st}$ June 2012 and 5$^{th}$ March 2007, respectively. Red and blue contours are swirls detected by ASDA \citep{Liu2019ApJ} with clockwise and anti-clockwise rotations, respectively. Panels b) and d) are the $\Gamma_2$ maps, corresponding to observations in panels a) and c). $\Gamma_2$ values are used to define edges of swirls \citep[see, e.g.,][]{Graftieaux2001, Liu2019ApJ}. Figure axes represent physical distances across the surface of the Sun (in Mm), with the origin of chosen domain placed at the centre of the observing FOV.}\label{fig:example}
\end{figure*}

Five sets of data are utilised in this study. The first three sets of data consist of wide-band photospheric images at Fe {\sc{i}} 630.25 nm, chromospheric images at the H$\alpha$ line core with a central wavelength of 656.3 nm, and chromospheric images at the Ca {\sc{ii}} line core with a central wavelength of 854.2 nm by SST/CRISP \citep{Scharmer2003,Scharmer2006,Scharmer2008} between 08:07:22 UT and 09:05:44 UT on 21$^{st}$ June 2012. The target was a quite-Sun region close to the disk centre ($x_c=-3$\arcsecs, $y_c=70$\arcsecs) with a FOV of 55\arcsecs$\times$55\arcsecs. The spatial and temporal resolutions are 0.1\arcsecs and 8.25~s, respectively. The black-white background in Figure~\ref{fig:example}a) depicts an example of the SST chromospheric images at the H$\alpha$ line core.

The other two sets of data consist of blue-continuum (FG-Blue) photospheric images with a central wavelength of 450.45 nm, and photospheric/chromospheric images at the Ca {\sc{ii}} H line with a central wavelength of 396.85~nm taken by Hinode/SOT \citep{Kosugi2007,Tsuneta2008} between 05:48:03 UT and 08:29:59 UT on the 5$^{th}$ March 2007. Each of them contains 1515 images of a quiet-Sun region close to the disk centre ($x_c$=5.3\arcsecs, $y_c$=4.1\arcsecs) with a field-of-view (FOV) of $\sim56$\arcsecs$\times28$\arcsecs. The spatial and temporal resolutions are 0.1\arcsecs and 6.42~s, respectively. Figure~\ref{fig:example}c) depicts an example of the SOT FG-blue photospheric images. We note that the broadband Ca {\sc ii} H observations by Hinode/SOT cover a wide range of altitudes from the photosphere to the chromosphere \citep{Rutten2004, Carlsson2007}.

The automated swirl detection algorithm ASDA \citep{Liu2019ApJ} was applied to every two successive images in each data set. ASDA contains two essential steps to perform the detection of swirls: 1) velocity field estimation using FLCT \citep{Welsch2004, Fisher2008} and 2) vortex identification using two parameters ($\Gamma_1$ and $\Gamma_2$) proposed by \cite{Graftieaux2001}. For each point in the velocity field, 49 points around it are used to calculate $\Gamma_1$ and $\Gamma_2$ for identifying the centres and edges of swirls, respectively. A detailed description of ASDA and how the parameters including location, radius, rotating speed, expanding/shrinking speed and lifetime of swirls are extracted from the $\Gamma_1$ and $\Gamma_2$ values could be found in \cite{Liu2019ApJ}. In summary, $\Gamma_2$ is used to identify swirl candidates and regions they cover, while $\Gamma_1$ is used to quantify swirl strength. Red and blue curves in Figure~\ref{fig:example}a) and c) are swirls detected by ASDA from the example SST H$\alpha$ chromospheric observation and SOT FG-blue photospheric observation, with clockwise rotations (red) and anti-clockwise rotations (blue), respectively. Figure~\ref{fig:example}b) and d) are the distributions of the $\Gamma_2$ values calculated from the example observations. Figure~\ref{fig:mosaic} depicts examples of individual swirls detected using ASDA from both SST and Hinode/SOT observations, with rows 2-6 displaying zoom-in views of the orange dashed boxes in the first row. Black and white backgrounds are the intensity observations and green streamlines are velocity fields estimated using FLCT. Red and blue contours are the edges of the example swirls with clockwise and counter-clockwise rotations, respectively. Panels in the second row are images and velocity fields $\sim$5 minutes before the example swirls, with curved streamlines at similar locations to where the swirls will form indicating their prerequisites. The third to the sixth rows are the first, middle and last frames of the example swirls, respectively. Consistent with the statistical findings in \cite{Liu2019ApJ} and \cite{Liu2019AA}, these swirls are located at intergranular lanes (dark features in the first, third, fourth and fifth columns). The last row shows the photospheric and chromospheric conditions $\sim$5 minutes after the example swirls.

To explore the relationship between photospheric and chromospheric swirls, bypassing a number of difficulties such as selection effects, the inclined magnetic field from the photosphere to the chromosphere, and the irregular shape of swirls, \cite{Liu2019NC} proposed a new method to calculate the correlation coefficient between photospheric and chromospheric swirls. In this work, to quantify the collective behaviour of swirls in the same data set (i.e., the same layer in the solar atmosphere), we have adapted the above method, as follows:

\begin{itemize}
\item For the $\Gamma_2$ maps of all images in a given layer, set all points to be 0 except those greater (less) than $2/\pi$ ($-2/\pi$), which are set to be 1 (-1).\\
\item Take the first $\Gamma_2$ map from the above as the reference and mark it as $\Gamma_{t1}$. Define $T_1$ as the sum of the absolute values of all points in $\Gamma_{t1}$.
\item Take the second $\Gamma_2$ map and mark it as $\Gamma_{t2}$. Define $T_2$ as the sum of the absolute values of all points in $\Gamma_{t2}$.
\item Multiply $\Gamma_{t1}$ and $\Gamma_{t2}$ point-by-point to obtain the correlation map $C$. Their correlation coefficient {$CC$} is then define as $(\sum{C}) / T$, where $T=max(T_1, T_2)$.\\
\item Repeat the above processes for the rest frames to obtain their correlation coefficients with the first frame of the data set. 
\end{itemize}

\begin{figure*}[ht!]
\includegraphics[width=\textwidth]{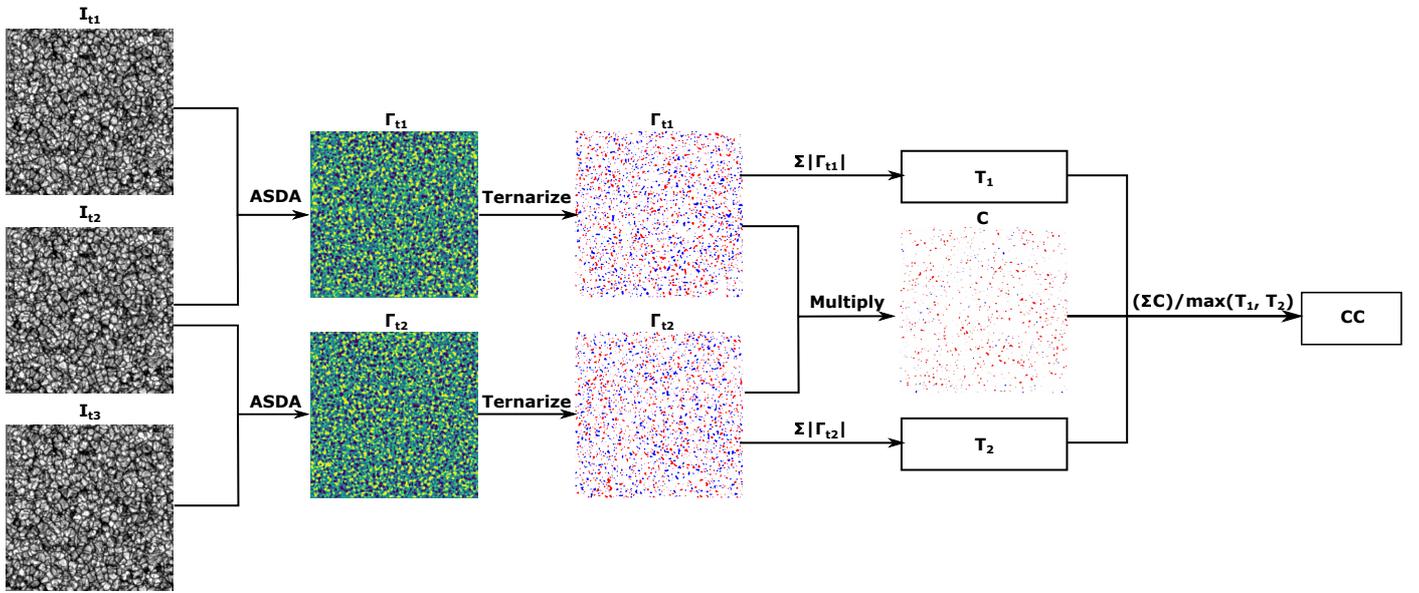}
\caption{\textbf{Flowchart of how the correlation coefficient (CC) is calculated}. $I_{t1}$, $I_{t2}$ and $I_{t3}$ are three frames of the intensity observations. Two $\Gamma_2$ maps ($\Gamma_{t1}$ and $\Gamma_{t2}$) are generated from these three intensity observations employing ASDA \citep{Liu2019ApJ}. They are further ternarised to contain only values of -1, 0 and 1. These ternarised $\Gamma_2$ maps are then used to calculate CC. For more information, see the text provided in Sect~\ref{sec:dam}.  \label{fig:flow}}
\end{figure*}

We can see from the definition of CC that it evaluates the similarity of swirl distribution between two frames. The above process is shown in Figure~\ref{fig:flow}.

\section{Results} \label{sec:result}

\subsection{Overall Parameters}
\begin{figure*}[ht!]
\includegraphics[width=0.9\textwidth]{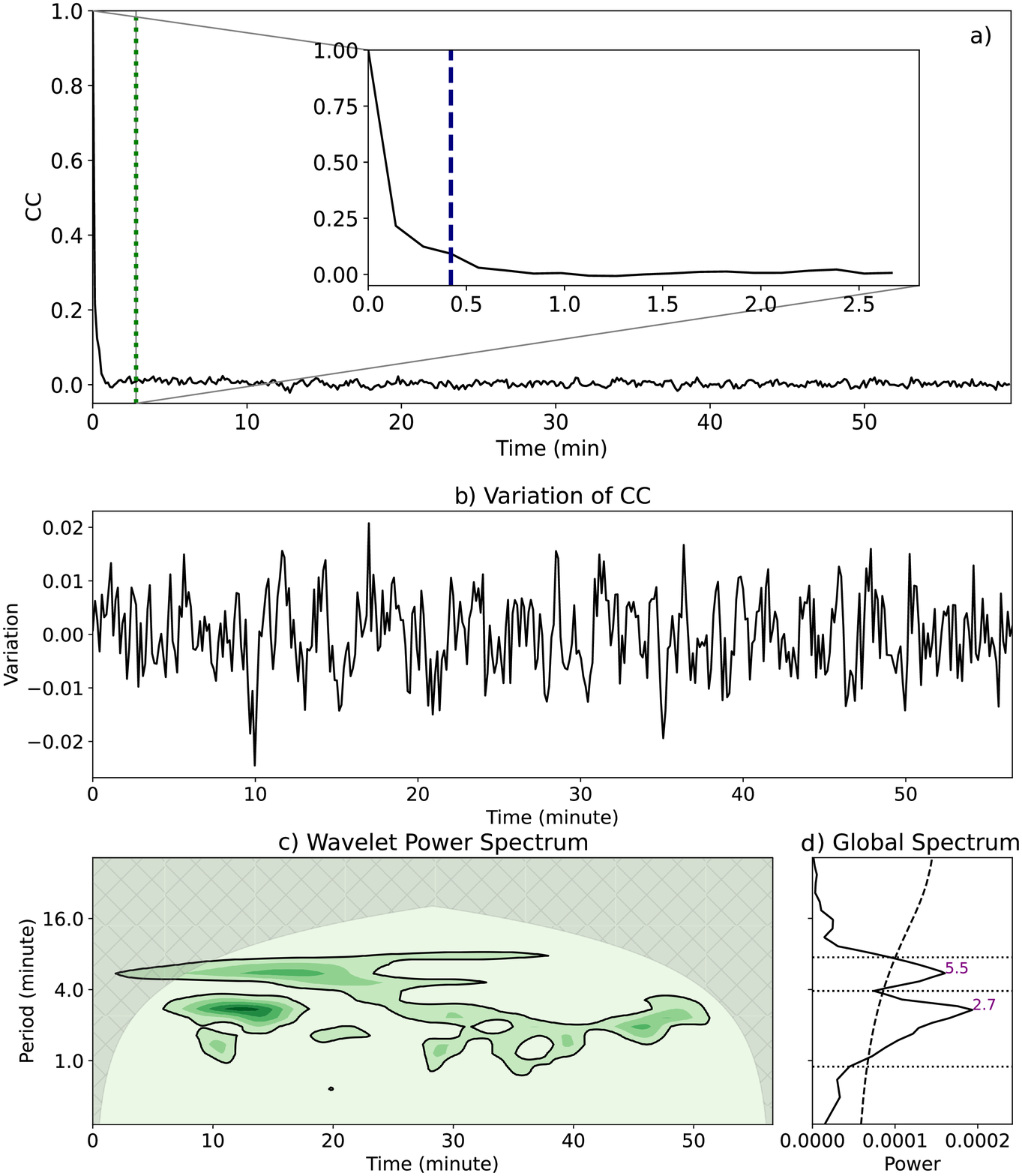}
\centering
\caption{\textbf{Variation of CC and its wavelet power spectrum}. Panel a) is the distribution of CC v.s. time for the SST H$\alpha$ line core chromospheric swirls. See Sect.~\ref{sec:dam} for the definition of CC. The inset in a) is a zoom-in view of the region before the green dashed line. The blue dashed line corresponds to a CC of 0.1. Panel b) shows the variation of CC after applying a highpass filter and subtracting its average value. Panel c) is the corresponding wavelet power spectrum with darker colours for higher powers. Black solid curves are the local 95\% confidence levels. The black solid curve in panel d) is the global wavelet power with purple text marking peaks above the 95\% confidence level (black dashed line). The black dotted lines are used to determine the extension of each peak.  \label{fig:cc}}
\end{figure*}

Figure~\ref{fig:cc}a) depicts the distribution of CC for SST H$\alpha$ line core chromospheric swirls. The correlation between each frame and the first frame drops quickly. The inset in panel a) shows a zoom-in view of the region before the green dashed line. The blue dashed line corresponds to a CC of 0.1 at $\sim$25~s. This is consistent with the average lifetime of photospheric and chromospheric swirls found using a different method \citep[$\sim16-23$~s,][]{Liu2019ApJ, Liu2019NC} and, again, confirms the short-lived characteristics of these small-scale structures.

Wavelet analysis \citep{Torrence1998} is employed to explore any potential periodicities in CC. To avoid the influence of the first few frames which have a high correlation with the first frame, all frames before the green dashed line in Figure~\ref{fig:cc}a) are omitted. Normally, an overall trend needs to be removed from the original time series to get rid of the unwanted periods introduced by the long-term variation. This overall trend is usually generated as the rolling average with a certain window width of the original time series. However, this needs very careful consideration because fake periods with values close to the window width can be introduced and thus contaminate the wavelet power spectrum.

Instead of subtracting the rolling average, we remove the long-term trend of the time series by applying a highpass filter to it that greatly weakens all signals with a period longer than $1/4$ of the length of the original time series. This approach is preferred as it does not introduce any fake periods. The processed time series is then used for the wavelet analysis after subtracting its average value. Figure~\ref{fig:cc}b) shows the processed CC. Its wavelet power spectrum is shown in Figure~\ref{fig:cc}c) with darker colours depicting stronger powers and black curves depicting 95\% confidence levels. Several periods from 2 min to 8 min can be found.

These periods are more obvious in the global wavelet spectrum (Figure~\ref{fig:cc}d). Two significant peaks at 2.7 min and 5.5 min are found above the 95\% confidence level (black dashed curve in Figure~\ref{fig:cc}d). Horizontal dotted lines are used to estimate the extension of each peak. For a given peak, its extension is defined by the distance between the local minima or the local 95\% confidence levels (whichever is closer to the target peak) on both sides. Periods found by the wavelet analysis on the CC values of swirls in the SST H$\alpha$ line core observations are then determined as 2.7$\pm$1.5 min and 5.5$\pm$1.8 min. Similar approaches utilizing the wavelet analysis have been applied to other properties of SST H$\alpha$ line core chromospheric swirls including the total number of swirls in each frame ($N$), their average intensity($\overline{I}$), average radius ($\overline{R}$), average rotating speed ($\overline{v_r}$) and average absolute expanding/shrinking speed ($\overline{v_e}$). Similar periods from 2 min to 13 min are found, which are shown as the fourth row in Table~\ref{tb_periods}. Note that any periods longer than $1/4$ of the length of the time series are crossed out in the table.

\begin{figure*}[ht!]
\centering
\includegraphics[width=0.9\textwidth]{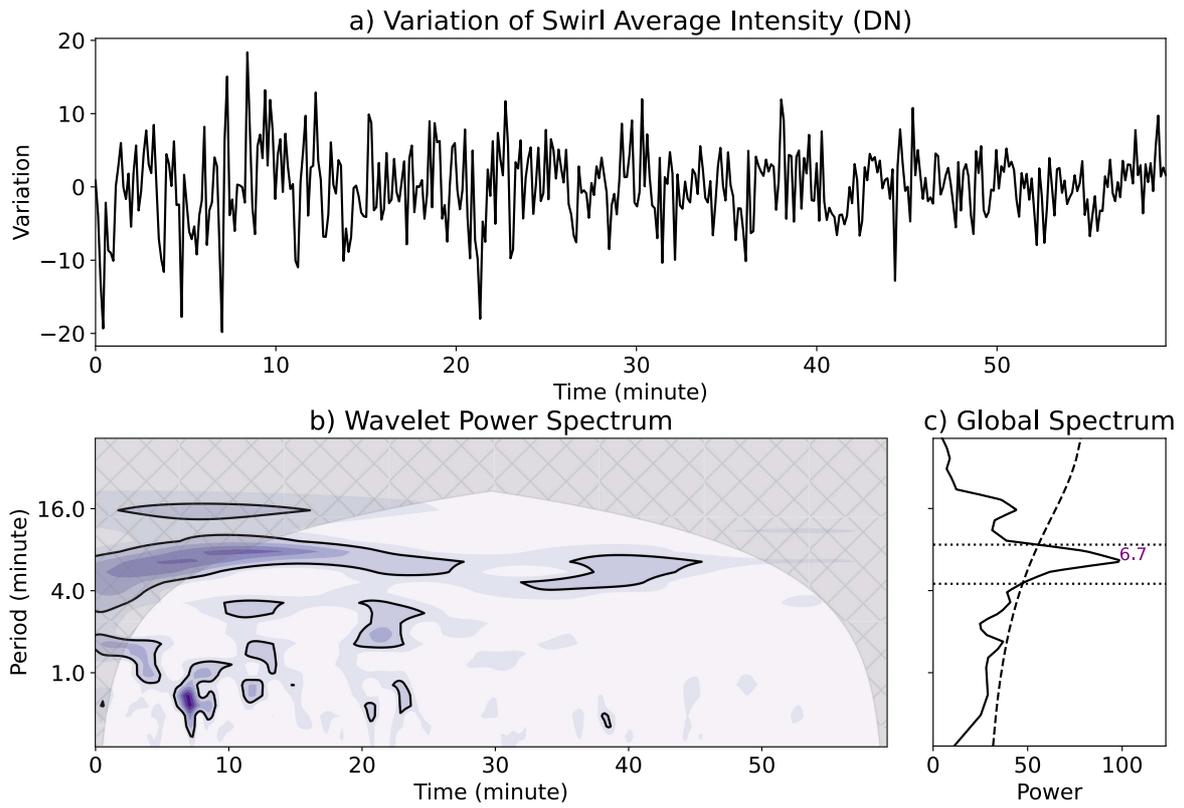}
\caption{\textbf{Variation of the average intensity of swirls and its wavelet spectrum}. Panels a) to c) are similar to panels b) to d) in Figure~\ref{fig:cc}, but for the SST Fe {\sc i} 6302 wide band photospheric swirls.  \label{fig:dn}}
\end{figure*}

Figure~\ref{fig:dn}a) shows the variation of the average intensity of SST Fe {\sc i} wideband photospheric swirls, after applying the highpass filter and subtracting its average value. Figure~\ref{fig:dn}b) and c) are the wavelet power spectrum and the global wavelet power, respectively. Only one significant period at 6.7$\pm$1.1 min above the 95\% confidence level has been found. Periods found from the time series of other parameters of the SST photospheric swirls range from 2 min to 13 min (second row in Table~\ref{tb_periods}). The rest of Table~\ref{tb_periods} show significant periods found for the six parameters ($CC$, $N$, $\overline{I}$, $\overline{R}$, $\overline{v_r}$ and $\overline{v_e}$) of swirls detected in the SOT FG-Blue, SOT Ca~{\sc ii} H and SST Ca~{\sc ii} line core. 84 significant periods have been found from all the 30 time series ranging from 1 min to 17 min, with an average value of 6.9$\pm$4.4 min. Out of all 84 periods, 45 ($\sim$54\%) are between 2 min to 9 min. 

\subsection{Periodicity of Individual Swirls}

\begin{figure*}[ht!]
\includegraphics[width=0.95\textwidth]{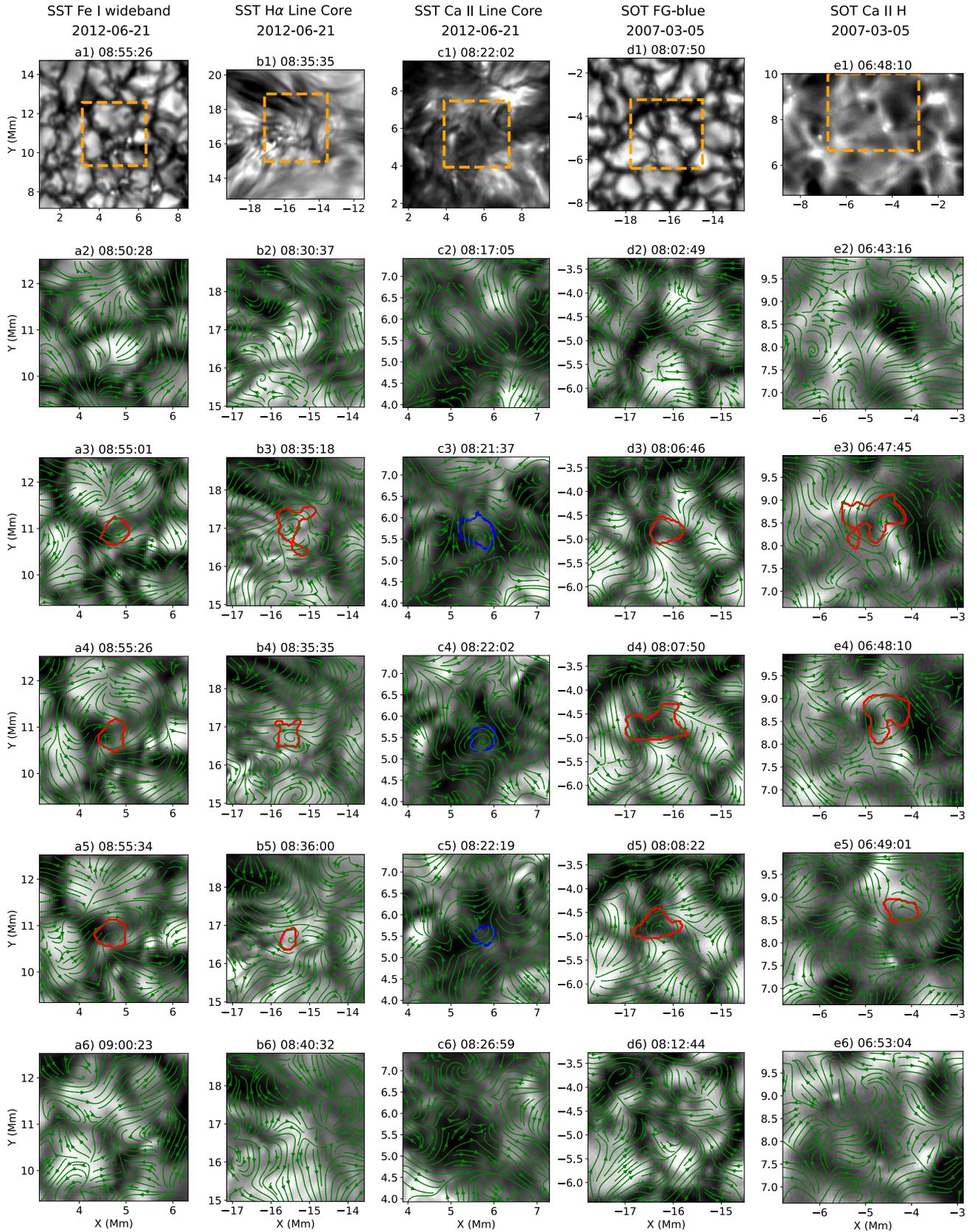}
\caption{\textbf{Examples of individual swirls in SST and Hinode Observations.} Black and white backgrounds are the corresponding intensities at each passband. Rows 2-6 are the zoom-in views of the orange boxes in the first row. Green arrows are velocity fields estimated using FLCT. Red and blue curves are the edges of the example swirls with clockwise and counter-clockwise rotations, respectively (see main text for details).} \label{fig:mosaic}
\end{figure*}

\begin{figure*}[ht!]
\centering
\includegraphics[height=0.9\textheight]{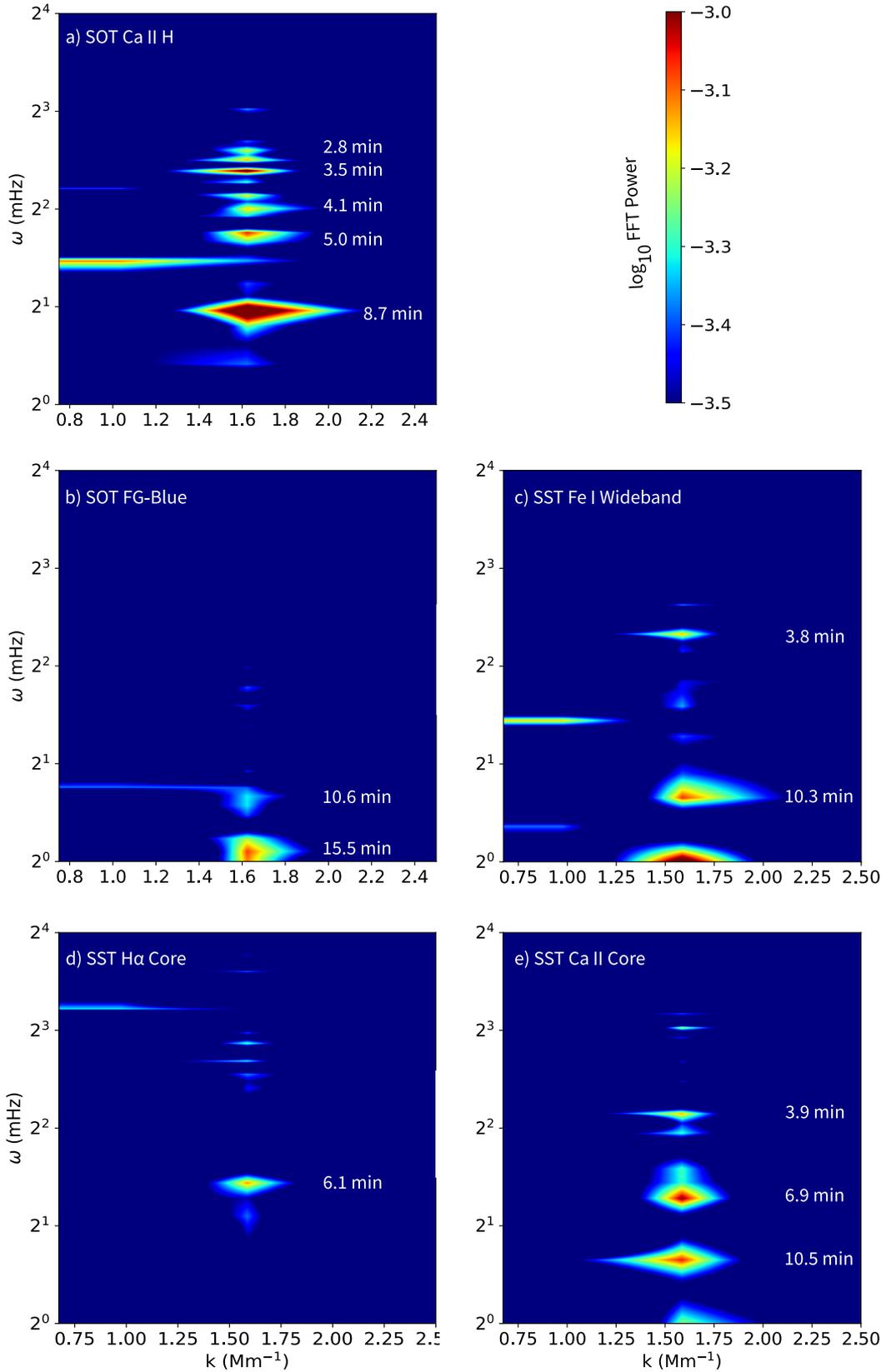}
\caption{\textbf{$k-\omega$ diagrams generated from the $\Gamma_2$ maps at the vicinity of the example swirls in different SOT and SST observations}. Numbers denote the corresponding central periods of the patches with high FFT powers.} \label{fig:fft}
\end{figure*}

To examine whether the above periodicities also exist in individual swirls, we further apply the wavelet analysis and Fast Fourier Transformation (FFT) to the $\Gamma_1$ and $\Gamma_2$ maps in small regions centred at the example swirls shown in Figure~\ref{fig:mosaic}. To conduct the FFT analysis, cogitating on the example swirl in the SOT Ca~{\sc ii} H observations (Fig.~\ref{fig:mosaic}e), a three-dimensional data cube $(x, y, t)$ $18\times18$ pix$^2$ around the centre of the swirl is extracted from its $\Gamma_2$ maps. $18\times18$ pix$^2$ is about twice of the average diameter of swirls detected, and chosen to avoid mixing signals of multiple swirls in the FOV. This data cube is then converted from the space-time domain $(x, y, t)$ to the wavenumber-frequency domain $(k, \omega)$ using FFT \citep[similarly applied by e.g., ][]{DeForest2004, LiuJ2012, 2017ApJ...842...59J}. We deduce from the definition of $\Gamma_2$ that it represents regions covered by swirl candidates, thus it indicates the appearance of swirls. $\Gamma_2$, instead of the originally observed intensity, is used in the FFT analysis because the former is directly related to swirls. As shown in Figure~\ref{fig:fft}a), several periods from $\sim3-8$~min can be identified as regions demonstrating high FFT power, where power is the square of the complex Fourier amplitudes. Applying similar approaches to the example swirls in other passbands reveals periods from $\sim3-16$~min (Fig.~\ref{fig:fft}), which are highly consistent with periods found from their corresponding overall parameters (Table~\ref{tb_periods}).

\begin{figure*}[ht!]
\centering
\includegraphics[width=0.9\textwidth]{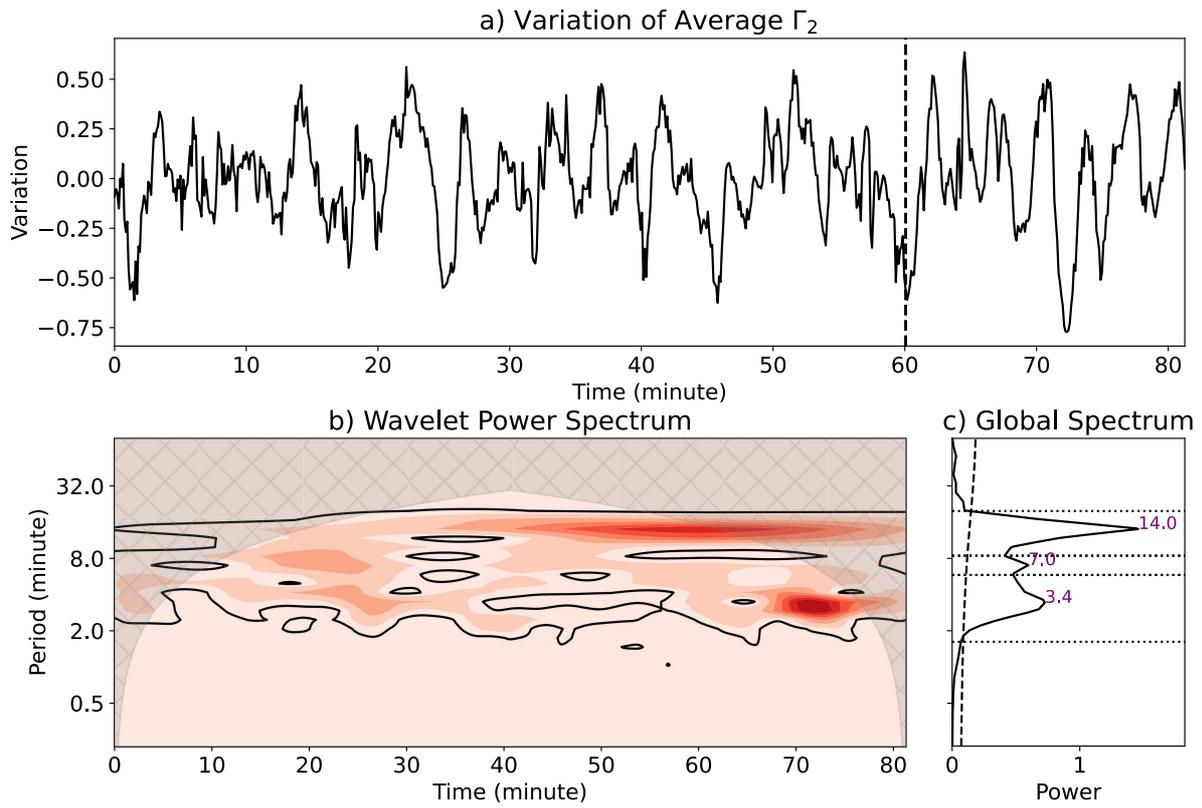}
\caption{\textbf{Variation of the average $\Gamma_2$ values in small regions centred at the SOT Ca~{\sc II} H example swirl and its wavelet spectrum}. Panels a) to c) are similar to panels b) to d) in Figure~\ref{fig:cc}, except that the vertical dashed line depicts the moment when the example swirl happens.} \label{fig:indiv}
\end{figure*}

The upper panel in Figure~\ref{fig:indiv} shows the variation of the average $\Gamma_2$ values in the area of $18\times18$ pix$^2$ around the centre of the SOT Ca~{\sc ii} H example swirl (Fig.~\ref{fig:mosaic}e), with the vertical dashed line denoting the moment of the example swirl in Figure~\ref{fig:mosaic}e4). Some periodicities can be easily pointed out by visual inspection. These periodicities are further confirmed by the wavelet analysis which shows significant wavelet powers above the 95\% confidence level at around $3.4$, $7.0$ and $14.0$ min. Applying the same approach to the average $\Gamma_1$ values around the SOT Ca~{\sc II} H example swirl reveals almost identical periods at around $3.6$, $7.0$ and $14.0$ min (first row, third column in Table~\ref{tb_indiv}).

Table~\ref{tb_indiv} lists all periods found from wavelet analysis of the average $\Gamma_1$ and $\Gamma_2$ values around the example swirls in the five studied passbands. Similarly as the SOT Ca~{\sc II} H example swirl, periods found from the $\Gamma_1$ values are almost identical to those from the $\Gamma_2$ values, and most of these periods fall within the range of $3$ to $14$ min. Taking that both $\Gamma_1$ and $\Gamma_2$ indicate the appearance of swirls, the above results suggest that, besides the collective behaviour of swirls, the occurrence of individual swirls also exhibits periods from around $3$ to $14$ min.

\subsection{Distribution of Periods}
\begin{figure*}[ht!]
\centering
\includegraphics[width=0.7\textwidth]{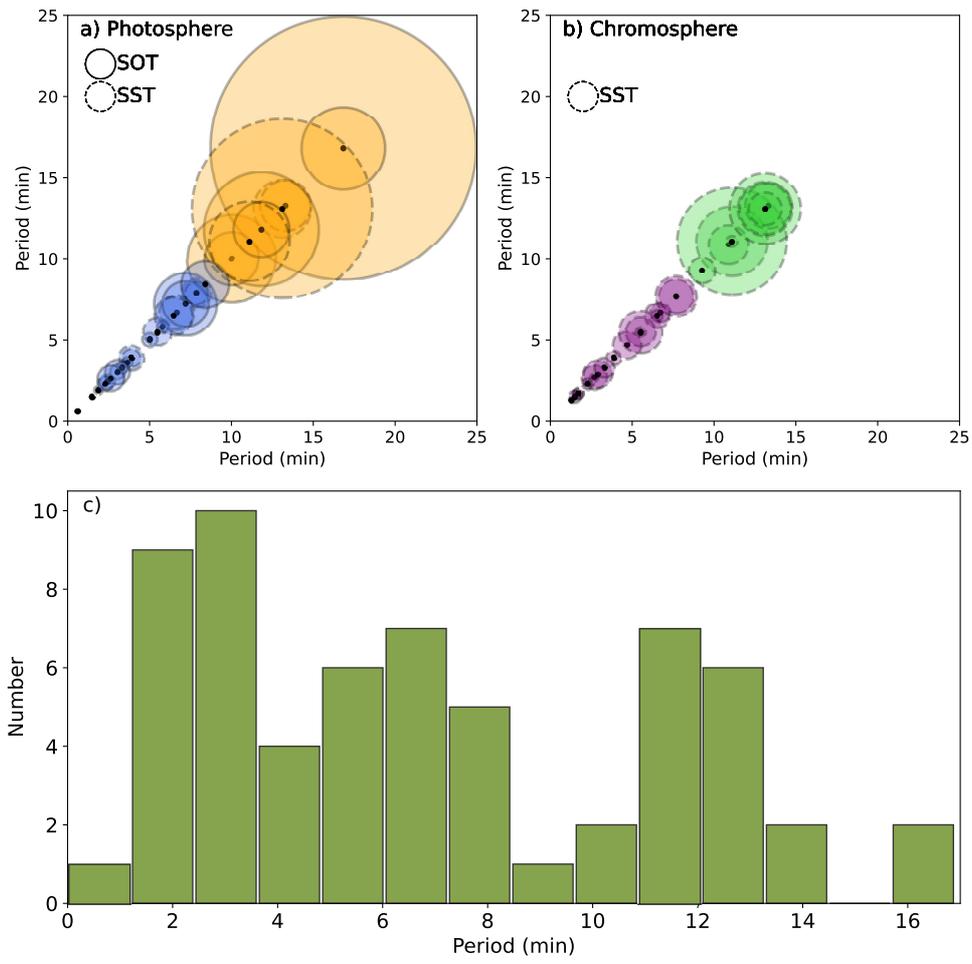}
\caption{\textbf{Distributions of significant periods.} Panels a) and b) are the distributions of periods determined from all photospheric and chromospheric swirls. Black solid (dashed) circles are for SOT (SST) swirls. Different colors denote different clusters identified by the k-means clustering algorithm (see Sect.~\ref{sec:result}). Panel c) is the histogram of all periods in all five data sets.   \label{fig:period}}
\end{figure*}

Figure~\ref{fig:period} shows the distribution of periods found in the overall parameters of photospheric swirls (panel a) and chromospheric swirls (panel b), with black dots denoting the periods and circles denoting their extensions. Swirls detected from the SOT Ca{\,}{\sc ii}{\,}H observations have been omitted from the above two panels (but included in panel c), as they might consist of both photospheric and chromospheric swirls due to the complicated formation heights of the SOT Ca{\,}{\sc ii}{\,}H broadband filter. Except that there are two above 15 min among the phototopheric periods due to the presence of Hinode/SOT observations, there is no obvious difference between photospheric and chromospheric periods, with their corresponding average values as 7.1$\pm$4.4 min and 6.7$\pm$4.1 min. Figure~\ref{fig:period}c) shows the distribution of all photospheric and chromospheric periods. It is shown again that most periods fall within the range of 3 min to 8 min. Meanwhile, another range of 10 min to 14 min, which is consistent with the second harmonics of the 3 min to 8 min periods, covers the second most periods. To further examine the above findings, we perform a machine learning method - the k-means clustering \citep{Macqueen1967}, which is one of the simplest but most effective algorithms for clustering purposes. It is worth noting that, k-means clustering is an unsupervised machine-learning algorithm which does not need the presence of labels. It takes a set of data as the input and automatically groups them into different clusters among which data points in each cluster have similar properties (i.e., periods in this case). 

The method relies on a given set of data containing $n$ points ($\mathbf{x_1}, \mathbf{x_2}, ..., \mathbf{x_n}$) to be divided into $k$ clusters ($S_1, S_2, ..., S_k$). The target of the k-means clustering is to separate all data into $k$ clusters of equal variance while minimizing the inertia defined as

\begin{equation}
    \sum_{i=0}^n min(||x_i-\mu_j||^2),
    \label{eq_kmeans}
\end{equation}

\noindent where $\mu_j$ are the centroids (average value) of each cluster. The k-clustering method is applied to the photospheric periods with a $k$ value of 2. The number of 2 is chosen because we expect 2 main period clusters as found above. Two distinct clusters (blue and orange circles in Figure~\ref{fig:period}a) are indeed found with average periods of 4.6$\pm$2.3 min and $12.8\pm$2.3 min, respectively. Application of the k-clustering method to the chromospheric periods reveals a very similar result. The two clusters (green and pink circles in Figure~\ref{fig:period}b) have average periods of 4.1$\pm$2.1 min and 11.9$\pm$1.3 min, respectively.

\section{Conclusions and Discussions} \label{sec:conc}

In this paper, we conducted a statistical study of oscillations related to photospheric and chromospheric swirls detected from five SST and SOT data sets using the automated swirl detection algorithm (ASDA). Periods with significant wavelet powers above their corresponding 95\% confidence levels were found from the time series of the correlation coefficient ($CC$), the number of swirls per frame ($N$), average intensity ($\overline{I}$), average radius ($\overline{R}$), average rotating speed ($\overline{v_r}$) and average expanding/shrinking speed ($\overline{v_e}$) of swirls detected in all five data sets. These periods range from 1 min to 17 min, with an average value of 6.9$\pm$4.4 min. Two clusters of periods have been found using the k-means clustering algorithm with one from 3 min to 8 min and the other from 10 min to 14 min, with the latter being approximately twice the former. More than half ($\sim$54\%) of these periods lie in or are very close to the 3 min to 8 min range. Similar periods have further been discovered from the $\Gamma_1$ and $\Gamma_2$ maps in the areas centred at five example swirls in the five different SST and SOT passbands using both the FFT and wavelet analysis.

The exact relation between the found $3-8$ min and $10-14$ min periods is unknown. Realistic numerical simulations might be needed to establish their physical relations. We also note that there are two periods less than $1$ min detected in the number of swirls ($N$) but not in other parameters (Table~\ref{tb_periods}). Moreover, the maximum periods found in SOT observations are overall larger than those in SST observations for the same parameters. It is unclear whether the above differences are physical or artificial. Further studies are needed to investigate whether these are due to the fact that the utilised SOT observations cover a longer time span.

Nevertheless, these periods found in swirls remind us of the well-known $3-8$ minute $p$-mode oscillations of the Sun \citep[e.g.,][]{Bahng1962, Ulrich1976, Zirker1980}, which are the result of the globally coherent acoustic waves with pressure as the restoring force. Table~\ref{tb_intensity} lists all significant periods found from the wavelet analysis of the average intensity across the whole field-of-view (FOV) of the SST and SOT observations. Most of the periods lie within the $5-8$~minute range (and $11-12$~minutes, which are twice that of the $5$-minute oscillations), providing evidence of the existence of global $p$-mode oscillations spanning the photosphere to the chromosphere. These periodicities also agree well with those found in the studied parameters of swirls detailed in this paper. Previously, oscillations with periods of the $p$-mode were also observed in a large variety of structures in the solar atmosphere from the photosphere to the corona, including but not limited to sunspots \citep[e.g.,][]{Cally2016}, coronal loops \citep[e.g.][]{Moortel2006}, plumes \citep[e.g.,][]{LiuJ2015}, and spicules \citep[e.g.,][]{Pontieu2004}.

Our results suggest that the global $p$-mode oscillations modulate not only the occurrence  manifested by periods in $CC$ and $N$, and periods in the investigated example swirls) but also the properties of photospheric swirls. Chromospheric swirls have also been found to possess almost identical periods. We note that employing simultaneous photospheric and chromospheric observations and 3D numerical simulations, \cite{Liu2019NC} found that photospheric swirls could trigger Alfv{\'e}n pulses which propagate upward into the chromosphere and lead to the observed chromospheric swirls. This suggests that the global $p$-modes could play a very important role in generating photospheric swirls thus Alfv{\'e}n pulses, chromospheric swirls and spicules, recalling that many studies have found that spicules could be driven by rotational motions at their footpoints \citep{Oxley2020, Scalisi2021, Scalisi2021b, Battaglia2021}. Given that most photospheric swirls appear to be located at intergranular lanes \citep[e.g.,][]{Wang1995, Bonet2008, Liu2019ApJ}, we suggest that they could be formed as a result of horizontal velocity flows \citep[e.g.,][]{Murawski2018} that are modulated by the global $p$-modes. However, the exact physical processes of how photospheric swirls are formed due to global $p$-modes and how the vertical oscillations in the $p$-modes are converted into the horizontal rotations in swirls are still unclear. This may be answered by investigating realistic 3D numerical simulations similar to the ones in \cite{Liu2019AA}, where photospheric swirls were detected in the simulated photosphere undergoing $p$-mode oscillations.

If this scenario will be proven, a new way for the global $p$-mode oscillation to channel both energy (via triggering Alfv{\'e}n pulses) and material (via triggering spicules) into the upper solar atmosphere will be established. Here, the $p$-mode oscillations themselves do not need to leak into the upper solar atmosphere in inclined magnetic field lines to trigger spicules as suggested by \cite{Pontieu2004}. Instead, they can also transport energy and mass into the upper atmosphere in less inclined magnetic lines via photospheric swirls.

\begin{acknowledgements}
Hinode is a Japanese mission developed and launched by ISAS/JAXA, with NAOJ as domestic partner and NASA and STFC (UK) as international partners. Hinode is operated by these agencies in co-operation with ESA and NSC (Norway). The Swedish $1$-m Solar Telescope is operated on the island of La Palma by the Institute for Solar Physics of Stockholm University in the Spanish Observatorio del Roque de los Muchachos of the Instituto de Astrof\'isica de Canarias. J.L. acknowledges the support from the National Key Technologies Research, Development Program of the Ministry of Science and Technology of China (2022YFF0711402) and the NSFC Distinguished Overseas Young Talents Program. D.B.J. and J.L. acknowledge support from the Leverhulme Trust via grant RPG-2019-371 and wish to thank the UK Space Agency for a National Space Technology Programme (NSTP) Technology for Space Science award (SSc~009). D.B.J. is also thankful to the UK Science and Technology Facilities Council (STFC) for grant Nos. ST/T00021X/1 and ST/X000923/1. R.E. is grateful to STFC for grant No. ST/M000826/1 and the Hungarian Scientific Research Fund (OTKA, grant No. K142987) of the National Research and Innovation Office (NKFIH), Hungary for the support received. R.E. also acknowledges the support received by the CAS Presidents International Fellowship Initiative grant No. 2019VMA052 and the Royal Society (IE161153).
\end{acknowledgements}

%
\bibliographystyle{aa} 

%

\clearpage

\begin{table*}[ht!]
\caption{Periods in units of minutes found from the wavelet analysis of the overall parameters of SOT and SST photospheric and chromospheric swirls.}
\label{tb_periods}
\begin{tabular}{p{2.5cm}|p{2cm}|p{2cm}|p{2cm}|p{2cm}|p{2cm}|p{2cm}}
Passband & CC & N & $\overline{I}$& $\overline{R}$&$\overline{v_r}$& $\overline{v_e}$ \\
\hline 
\hline
SOT FG-Blue & $\begin{tabular}{@{}c@{}}10.0$\pm$2.7 \\ 16.8$\pm$8.1 \\ \end{tabular}$ & $\begin{tabular}{@{}c@{}}0.6$\pm$0.1 \\ 5.8$\pm$0.3 \\ 10.0$\pm$1.7 \\ \sout{33.6$\pm$3.4} \\ \end{tabular}$ & $\begin{tabular}{@{}c@{}}2.4$\pm$0.5 \\ 3.6$\pm$0.3 \\ 7.2$\pm$1.9 \\ \sout{33.6$\pm$7.9} \\ \end{tabular}$ & $\begin{tabular}{@{}c@{}}2.6$\pm$0.8 \\ 8.4$\pm$1.4 \\ 11.8$\pm$1.7 \\ \sout{33.6$\pm$2.5} \\ \end{tabular}$ & $\begin{tabular}{@{}c@{}}3.0$\pm$0.4 \\ 7.2$\pm$1.4 \\ 16.8$\pm$2.5 \\ \sout{28.2$\pm$5.0} \\ \sout{47.4$\pm$3.6} \\ \end{tabular}$ & $\begin{tabular}{@{}c@{}}3.0$\pm$0.8 \\ 5.0$\pm$0.4 \\ 11.8$\pm$3.5 \\ \sout{28.2$\pm$5.0} \\ \sout{47.4$\pm$2.4} \\ \end{tabular}$ \\ 
 \hline 
SST Fe {\sc{i}} Wideband & $\begin{tabular}{@{}c@{}}3.9$\pm$0.7 \\ 7.9$\pm$0.9 \\ 13.1$\pm$1.7 \\ \end{tabular}$ & $\begin{tabular}{@{}c@{}}1.5$\pm$0.1 \\ 3.3$\pm$0.3 \\ \end{tabular}$ & 6.7$\pm$1.1 & $\begin{tabular}{@{}c@{}}2.3$\pm$0.5 \\ 6.5$\pm$1.1 \\ 11.1$\pm$2.5 \\ \end{tabular}$ & $\begin{tabular}{@{}c@{}}1.9$\pm$0.3 \\ 3.9$\pm$0.5 \\ 13.3$\pm$1.5 \\ \end{tabular}$ & $\begin{tabular}{@{}c@{}}5.5$\pm$0.8 \\ 7.9$\pm$0.7 \\ 13.1$\pm$5.5 \\ \sout{37.1$\pm$3.9} \\ \end{tabular}$ \\ 
 \hline 
SOT Ca {\sc{ii}} H & $\begin{tabular}{@{}c@{}}2.6$\pm$0.6 \\ 5.0$\pm$0.9 \\ 14.2$\pm$3.6 \\ \end{tabular}$ & $\begin{tabular}{@{}c@{}}0.6$\pm$0.1 \\ 4.2$\pm$0.7 \\ 7.2$\pm$0.9 \\ 14.2$\pm$0.9 \\ \end{tabular}$ & $\begin{tabular}{@{}c@{}}5.0$\pm$1.6 \\ 9.8$\pm$0.8 \\ 14.2$\pm$0.8 \\ \end{tabular}$ & $\begin{tabular}{@{}c@{}}1.6$\pm$0.5 \\ 3.6$\pm$0.4 \\ 6.0$\pm$1.1 \\ 10.0$\pm$0.8 \\ 16.8$\pm$1.1 \\ \end{tabular}$ & $\begin{tabular}{@{}c@{}}1.8$\pm$0.6 \\ 5.0$\pm$0.9 \\ 10.0$\pm$1.4 \\ \end{tabular}$ & $\begin{tabular}{@{}c@{}}1.6$\pm$0.3 \\ 3.0$\pm$0.2 \\ 5.0$\pm$0.9 \\ 10.0$\pm$1.1 \\ \end{tabular}$ \\ 
 \hline 
SST H$\alpha$ Core & $\begin{tabular}{@{}c@{}}2.7$\pm$0.8 \\ 5.5$\pm$0.9 \\ \end{tabular}$ & $\begin{tabular}{@{}c@{}}2.7$\pm$0.3 \\ 6.7$\pm$0.6 \\ 10.9$\pm$1.2 \\ \end{tabular}$ & $\begin{tabular}{@{}c@{}}7.7$\pm$1.1 \\ 11.1$\pm$0.3 \\ \end{tabular}$ & $\begin{tabular}{@{}c@{}}2.3$\pm$0.4 \\ 5.5$\pm$1.3 \\ 9.3$\pm$0.8 \\ 13.1$\pm$1.0 \\ \end{tabular}$ & $\begin{tabular}{@{}c@{}}2.9$\pm$0.9 \\ 13.1$\pm$1.7 \\ \end{tabular}$ & $\begin{tabular}{@{}c@{}}1.7$\pm$0.3 \\ 3.3$\pm$0.6 \\ 6.5$\pm$0.8 \\ 13.3$\pm$1.4 \\ \end{tabular}$ \\ 
 \hline 
SST Ca {\sc{ii}} Core & $\begin{tabular}{@{}c@{}}1.7$\pm$0.3 \\ 3.9$\pm$0.4 \\ 11.1$\pm$2.2 \\ \end{tabular}$ & 5.5$\pm$0.3 & $\begin{tabular}{@{}c@{}}4.7$\pm$0.8 \\ 11.1$\pm$3.3 \\ \end{tabular}$ & $\begin{tabular}{@{}c@{}}1.3$\pm$0.2 \\ 2.3$\pm$0.2 \\ 13.1$\pm$2.2 \\ \end{tabular}$ & $\begin{tabular}{@{}c@{}}1.5$\pm$0.4 \\ 6.5$\pm$0.4 \\ \end{tabular}$ & $\begin{tabular}{@{}c@{}}7.7$\pm$1.3 \\ 13.1$\pm$1.6 \\ \end{tabular}$ \\ 
 \hline 
\multicolumn{7}{l}{\footnotesize Note---$CC$ is the correlation coefficient as defined in Sect.~\ref{sec:dam}. For swirls detected in each frame of observations, $N$ represents for the total} \\
\multicolumn{7}{l}{\footnotesize  number, $\overline{I}$ the average intensity, $\overline{R}$ the average radius, $\overline{v_r}$ the average rotating speed and $\overline{v_e}$ the average expanding/shrinking speed.}
\end{tabular}
\end{table*}

\begin{table*}[hbt!]
\caption{Periods in units of minutes found from the wavelet analysis at the vicinity of SOT and SST example swirls.}
\label{tb_indiv}
\begin{tabular}{p{2cm}|p{3cm}|p{3cm}|p{2cm}|p{2cm}|p{2.5cm}}
Passband & SOT FG-Blue & SST Fe \sc{i} Wideband & SOT Ca \sc{ii} H & SST H$\alpha$ Core & SST Ca \sc{ii} Core \\
\hline
\hline 
$\Gamma_1$& $\begin{tabular}{@{}c@{}}10.0$\pm$3.0 \\ 20.0$\pm$2.5 \\ \sout{28.2$\pm$5.3} \\ \end{tabular}$ & $\begin{tabular}{@{}c@{}}3.7$\pm$0.6 \\ 7.7$\pm$1.6 \\ \end{tabular}$ & $\begin{tabular}{@{}c@{}}3.6$\pm$0.6 \\ 7.0$\pm$1.4 \\ 14.0$\pm$2.6 \\ \end{tabular}$ & $\begin{tabular}{@{}c@{}}2.7$\pm$0.7 \\ 9.1$\pm$2.0 \\ \end{tabular}$ & $\begin{tabular}{@{}c@{}}1.9$\pm$0.3 \\ 3.3$\pm$0.4 \\ 5.5$\pm$0.9 \\ 10.9$\pm$0.7 \\ \end{tabular}$\\ 
\hline
 $\Gamma_2$& $\begin{tabular}{@{}c@{}}3.6$\pm$0.6 \\ 12.0$\pm$3.1 \\ 20.0$\pm$1.8 \\ \sout{28.2$\pm$5.7} \\ \end{tabular}$ & 6.5$\pm$2.2 & $\begin{tabular}{@{}c@{}}3.4$\pm$1.1 \\ 7.0$\pm$0.6 \\ 14.0$\pm$2.9 \\ \end{tabular}$ & $\begin{tabular}{@{}c@{}}2.7$\pm$0.6 \\ 6.3$\pm$1.8 \\ \end{tabular}$ & $\begin{tabular}{@{}c@{}}1.9$\pm$0.3 \\ 2.7$\pm$0.4 \\ 5.5$\pm$1.0 \\ 10.9$\pm$1.4 \\ \end{tabular}$\\ 
\hline 
\end{tabular}
\end{table*}

\begin{table*}[htb!]
\caption{Periods in units of minutes found from the wavelet analysis of average intensity throughout the entire FOV of different SOT and SST observations.}
\label{tb_intensity}
\begin{tabular}{p{2cm}|p{2cm}|p{3cm}|p{2cm}|p{2cm}|p{2.5cm}}
Passband & SOT FG-Blue & SST Fe {\sc{i}} Wideband & SOT Ca {\sc{ii}} H & SST H$\alpha$ Core & SST Ca {\sc{ii}} Core \\
\hline
\hline 
Periods & $\begin{tabular}{@{}c@{}}5.0$\pm$0.8 \\ 8.4$\pm$2.7 \\ \sout{28.2$\pm$7.4} \\ \end{tabular}$ & 5.5$\pm$3.1 & $\begin{tabular}{@{}c@{}}5.0$\pm$1.4 \\ 11.8$\pm$1.4 \\ \end{tabular}$ & $\begin{tabular}{@{}c@{}}6.5$\pm$1.5 \\ 11.1$\pm$2.2 \\ \end{tabular}$ & 4.7$\pm$1.4\\ 
\hline 
\end{tabular}
\end{table*}

\end{document}